%% file: Lat19.tex
\documentclass{PoS_mod}

\usepackage{amsmath,subfigure,latexsym,amssymb,cite}

\newcommand{\be}{\begin{equation}}
\newcommand{\ee}{\end{equation}}
\newcommand{\nn}{\nonumber}
\newcommand{\bea}{\begin{eqnarray}}
\newcommand{\eea}{\end{eqnarray}} 

\newcommand{\la}{\langle}
\newcommand{\ra}{\rangle}

\newcommand{\R}{{\kern+.25em\sf{R}\kern-.78em\sf{I} \kern+.78em\kern-.25em}}
\newcommand{\RR}{{\kern+.25em\sf{R}\kern-.6em\sf{I} \kern+.6em\kern-.25em}}
\newcommand{\N}{{\kern+.25em\sf{N}\kern-.78em\sf{I} \kern+.78em\kern-.25em}}
\newcommand{\C}{\mathbb{C}}
\newcommand{\Z}{\mathbb{Z}}

\newcommand{\CP}{\C {\rm P}}

\makeatletter
\@addtoreset{equation}{section}
\makeatother

\title{Meron- and Semi-Vortex-Clusters as Physical Carriers of Topological 
Charge and Vorticity\thanks{This work was supported by UNAM-DGAPA-PAPIIT, grant 
number IG100219, by the Albert Einstein Center for Theoretical Physics and by 
the European Research Council under the European Union's Seventh Framework 
Programme (FP7/2007-2013)/ERC grant agreement 339220, and by the Schweizerischer
Nationalfonds.}}

\ShortTitle{Clusters as Carriers of Topological Charge}

\speakershort{Wolfgang Bietenholz and Jo\~{a}o C. Pinto Barros}

\author{\speaker{Wolfgang Bietenholz}$^{\rm \,\, a}$,
        \speaker{Jo\~{a}o C. Pinto Barros}$^{\rm \,\, b}$, 
        Stephan Caspar$^{\rm \, c}$,\newline
        Manes Hornung$^{\rm \, b}$, and Uwe-Jens Wiese$^{\rm \, b}$\, \footnote{Combined contribution of both speakers.}
        \ \\
\vspace*{3mm}
\ \\
\ $^{\rm \ a}$ Instituto de Ciencias Nucleares, 
Universidad Nacional Aut\'{o}noma de M\'{e}xico (UNAM), \\
~~~~A.P.\ 70-543, C.P.\ 04510 Ciudad de M\'{e}xico, Mexico \vspace*{1mm} \\
\ $^{\rm \, b}$ Albert Einstein Center for Fundamental Physics,
Institute for Theoretical Physics, \\
~~~~University of Bern, Sidlerstrasse 5, CH-3012 Bern, Switzerland
\vspace*{1mm} \\
\ $^{\rm \, c}$ Institute  for  Nuclear  Theory,  University  of  Washington,  
Seattle,  WA  98195,  USA
\vspace*{3mm}

E-mail: \email{wolbi@nucleares.unam.mx, jpintobarros@itp.unibe.ch} \\ }

\abstract{In O($N$) non-linear $\sigma$-models on the lattice, the Wolff cluster
algorithm is based on rewriting the functional integral in terms of mutually
independent clusters. Through improved estimators, the clusters are directly
related to physical observables. In the $(N-1)$-d O($N$) model
(with an appropriately constrained action) the clusters carry an integer or
half-integer topological charge. Clusters with topological charge $\pm 1/2$
are denoted as merons. Similarly, in the 2-d O(2) model the clusters carry 
pairs of semi-vortices and semi-anti-vortices (with vorticity $\pm 1/2$) at 
their boundary. Using improved estimators, meron- and semi-vortex-clusters 
provide analytic insight into the topological features of the dynamics. We 
show that the histograms of the cluster-size distributions scale in the 
continuum limit, with a fractal dimension $D$, which suggests that the clusters 
are physical objects. We demonstrate this property analytically for merons and 
non-merons in the 1-d O(2) model (where $D=1$), and numerically for the 2-d 
O(2), 2-d O(3), and 3-d O(4) model, for which we observe fractal dimensions 
$D < d$. In the vicinity of a critical point, a scaling law relates $D$ to a
combination of critical exponents. In the 2-d O(3) model, meron- and 
multi-meron-clusters are responsible for a logarithmic ultraviolet divergence 
of the topological susceptibility.}

\FullConference{37th International Symposium on Lattice Field Theory
  - Lattice2019\\
		16-22 June 2019\\
		Wuhan, China }

\begin{document}

\section{Outline}

\input{outline}

\section{Lattice O($N$) models, topology, and 
meron-clusters}

\input{LatON}
 
\section{Cluster-size scaling for the 1-d O(2) model}

\input{1dO2}

\section{Fractal cluster dimension and the role of 
merons in the 2-d O(3) model}

\input{2dO3}

\section{Fractal dimension and vorticity in the 
2-d O(2) model}

\input{2dO2}

\section{The 3-d O(4) model and a scaling law}

\input{3dO4}

\section{Conclusions}

\input{conclu}

\section*{Acknowledgments}

\noindent We thank Martin L\"uscher for many interesting discussions. We also acknowledge
helpful correspondences with Ian Affleck, John Cardy, Alexander M.\ Polyakov, and
Alexander Zamolodchikov.

\end{document}

%% file: outline.tex
The functional integral of lattice O($N$) models can be expressed analytically
in terms of Wolff clusters, whose numerical simulation provides a most 
efficient algorithm \cite{Wolff1}. Here we are going to show that the clusters 
are physical objects whose 
size-distribution exhibits continuum scaling, with some fractal dimension 
$D \leq d$, where $d$ is the space-time dimension. Like instantons, clusters 
appear in Euclidean field configurations and are thus not directly accessible 
in physical experiments. However, unlike instantons, the clusters are not just
semi-classical objects, but determine the full lattice functional integral 
without any approximation. In models with topological sectors, {\it i.e.}\ for
$d=N-1$, the clusters turn out to be the physical carriers of integer or
half-integer topological charge. Clusters with topological charge 
$Q_{\cal C} = \pm 1/2$ are called {\em merons}, while those with $Q_{\cal C} = 0$ are 
{\em non-merons}, and we denote clusters with $|Q_{\cal C}| \geq 1$ as {\em multi-merons}. Via 
improved estimators, this classification provides both analytic insights and 
accurate numerical results for the topology-driven aspects of the dynamics.

Similarly, in the 2-d O(2) model the clusters carry a number of 
semi-vortex--semi-anti-vortex pairs at their boundaries. In this way, the 
cluster dynamics has the potential to provide a refined interpretation of the 
Berezinski\u{\i}-Kosterlitz-Thouless phase transition \cite{BKT1,BKT2}.

%% file: LatON.tex
Lattice O($N$) non-linear $\sigma$-models are formulated with classical spin 
variables $\vec e_x \in S^{N-1}$ ({\it i.e.}\ $\vec e_x \in \R^N$ and 
$|\vec e_x| = 1$) associated with the lattice sites $x$. We use both hypercubic 
and triangular space-time lattices with unit lattice spacing, and we always 
assume periodic boundary conditions in all directions.

In the numerical simulations we are going to use three lattice actions, each
being a sum of nearest-neighbor contributions, 
$S[\vec e \, ] = \sum_{\la xy \ra} s(\vec e_x,\vec e_y)$, with
\bea
s_{\rm standard}(\vec e_x,\vec e_y)&=&
\frac{1}{g^2} (1 - \vec e_x \cdot \vec e_y) \nn \\
s_{\rm topological}(\vec e_x,\vec e_y)&=&
\left\{ \begin{array}{ccccc}
0 &&&& {\rm if} ~~ \vec e_x \cdot \vec e_y > \cos\delta \\
+ \infty &&&& {\rm otherwise} \end{array} \right. \nn \\
s_{\rm constraint}(\vec e_x,\vec e_y)&=&
\left\{ \begin{array}{ccccc} 
\frac{1}{g^2} (1 - \vec e_x \cdot \vec e_y) &&&& 
{\rm if} ~~ \vec e_x \cdot \vec e_y > \cos\delta \\
+ \infty &&&& {\rm otherwise.} \end{array} \right.
\label{latactions}
\eea
The {\em topological lattice action} \cite{topact} is invariant under small 
deformations of the spin configuration. It belongs to the same universality 
class as the standard action, despite the fact that it does not have the 
correct classical continuum limit. Decreasing the angle $\delta$, which 
constrains the relative angle between nearest-neighbor spins, has the same 
effect as decreasing $g^2$. The constraint action modifies the standard 
action by a constraint angle. Optimized constraint actions have been used to 
almost completely eliminate lattice artifacts \cite{Bogli,optact}.

The most efficient known algorithm for the simulation of O($N$) models
is the Wolff cluster algorithm \cite{Wolff1}, which generalizes the 
Swendsen-Wang algorithm \cite{SwendsenWang} for the Ising model as follows:
{\small
\begin{enumerate}

\item Choose a random unit-vector $\vec r \in S^{N-1}$. A {\em spin flip}
is defined as the reflection at the hyperplane perpendicular to $\vec r$. It 
transforms a spin variable as
$
\vec e_x \to \vec e_x{\, '} = \vec e_x - 2 (\vec e_x \cdot \vec r) \, \vec r \ .
$

\item Consider any pair of nearest-neighbor spins, $\vec e_x$ and $\vec e_y$. 
Flipping one of these spins turns their contribution to the action 
$s(\vec e_x,\vec e_y)$ into a modified contribution 
$s(\vec e_x{\, '},\vec e_y) = s(\vec e_x,\vec e_y{\, '})$, and we define 
$\Delta s_{\la xy \ra} = s(\vec e_x{\, '},\vec e_y) - s(\vec e_x,\vec e_y)$. Now 
we set a {\em bond} between these two spins with probability
\be
p_{\la xy \ra} = \left\{ \begin{array}{ccccl}
0 &&& {\rm if} & \quad \Delta s_{\la xy \ra} \leq 0 \\
1 - \exp(- \Delta s_{\la xy \ra}) &&&  {\rm if} & \quad \Delta s_{\la xy \ra} > 
0 \ . \end{array} \right.
\label{bondprob}
\ee

\item A set of spins connected by bonds defines a {\em cluster.} In the
multi-cluster variant of the algorithm, which we use in this study, all spins 
in a cluster are flipped collectively with probability $1/2$. The
algorithm obeys detailed balance without any need for an accept/reject step.

\item Measure observables, if possible using an improved estimator (see below), 
and return to 1.

\end{enumerate}
}

This algorithm is far superior to local update algorithms, in particular close 
to criticality. The collective spin updates strongly reduce auto-correlations, 
and therefore almost completely eliminate critical slowing down \cite{Wolff1}.

Since the clusters can be flipped independently, a configuration with 
$n_{\cal C}$ clusters can be viewed as a member of a sub-ensemble of 
$2^{n_{\cal C}}$ configurations. For several observables of physical interest it 
is possible to average over this sub-ensemble analytically, thus increasing the 
statistics by a large amount without actually generating these configurations. 
This is known as an {\em improved estimator}.

Here we are going to elaborate on the physical interpretation of these 
clusters, cf.\ Section 1. In a continuum O($N$) model with $d=N-1$, the 
space of configurations decomposes 
into topological sectors. This means that each configuration $[\vec e \, ]$ 
(with finite action $S[\vec e \, ]$) carries a topological charge 
$Q[\vec e \, ] \in \Z$, which cannot be altered by continuously deforming the 
configuration (while keeping the action finite). On the lattice, usually 
all configurations can be continuously deformed into one another, such that the 
topological charge of a lattice configuration requires an appropriate 
definition. For the topological or constraint actions, infinite-action barriers
even exist on the lattice.

Here we apply the {\em geometric definition} \cite{BergLuscher}, which assigns 
a topological charge $Q[\vec e \, ] \in \Z$ to each lattice configuration (up 
to a subset of measure zero, unless there are sufficiently stringent infinite 
lattice action barriers). For a hypercubic lattice, a unit hypercube is divided 
into simplices with $N$ sites ({\it e.g.}\ for $d=2$ a plaquette is split into 
two triangles). The $N$ spins attached to a simplex represent $N$ points 
$\vec e_x$ which define a minimal spherical simplex in $S^{N-1}$ by interpolation along shortest geodesics. The oriented volume of 
this spherical simplex (with either a positive or negative sign), divided by the 
volume of $S^{N-1}$, defines the topological charge density 
$q_i \in \left[-\frac{1}{2},\frac{1}{2}\right]$ 
that is associated with the simplex labeled with $i$. For example, for $d=2$ a 
lattice triangle is mapped to a spherical triangle via the three spins 
$\vec e_x \in S^2$ at its vertices. The oriented area of the spherical triangle,
divided by $4 \pi$, then determines $q_i$. Due to periodic
boundary conditions, by construction the total topological charge of the
$(N-1)$-dimensional O($N$) model, $Q = \sum_i q_i \in \Z$, then measures the 
number of times the spin configuration covers the sphere $S^{N-1}$. This
geometric topological charge can be used, {\it e.g.}\ in the 1-d O(2), 2-d O(3),
and 3-d O(4) model. 

Since the topological charge density $q_i$ changes sign under cluster flip, we 
define the topological charge of a Wolff cluster ${\cal C}$ as \cite{meron}
\be \label{Qclust}
Q_{\cal C} = \tfrac{1}{2} (Q [\vec e \, ] - Q [\vec e \, ']) \in
\{ 0, \, \pm \tfrac{1}{2}, \, \pm 1, \, \pm \tfrac{3}{2}, \, \pm 2, \,
\dots \} \ .
\ee
Here $[\vec e \, ']$ is the configuration that one obtains after flipping the 
cluster ${\cal C}$, keeping all the other clusters fixed. In order to ensure 
that $Q_{\cal C}$, defined in this way, is independent of the relative 
orientations of all the other clusters, we impose a constraint angle $\delta$ 
between all nearest-neighbor spins that belong to one simplex, such that
\be \label{deltaconstraint}
\vec e_x \cdot \vec e_y > \cos\delta \geq - \frac{1}{N-1} \ . 
\ee
This implies that in the 1-d O(2) model no such constraint is necessary. For 
the 2-d O(3) model on the triangular lattice we use the constraint angle 
$\delta = 2 \pi/3$ such that $\cos\delta = - 1/2$, as first realized in 
Ref.\ \cite{meron}. On a square lattice one can also use $\cos\delta = 0$ for
nearest-neighbor spins, without constraining the relative angle between diagonal
next-to-nearest neighbors. Under the condition (\ref{deltaconstraint}), each 
cluster carries a uniquely defined topological charge $Q_{\cal C}$ and thus 
qualifies as a physical topological object that is independent of the other 
clusters. In that case the total topological charge is obtained as the sum of 
all cluster contributions, $Q = \sum_{\cal C} Q_{\cal C}$.

It is worth noting that, by construction (at least for the actions of 
eq.\ (\ref{latactions})) all spins within a given cluster are on the same side of
the reflection hyperplane. This feature is vital for the efficiency of the 
algorithm, because it prevents clusters from growing to unphysically large 
size. By flipping all clusters to the same side of the reflection hyperplane, 
one reaches a {\em reference configuration} which cannot cover $S^{N-1}$ and 
thus has 
vanishing topological charge. When we define the cluster charge $Q_{\cal C}$ by
a flip with respect to a reference configuration, one can distinguish merons
with $Q_{\cal C} = 1/2$ from anti-merons with $Q_{\cal C} = - 1/2$ in a meaningful
way. In the present context this distinction is not important and hence we
refer to all clusters with $|Q_{\cal C}| = 1/2$ as merons. Similarly, we denote
clusters with $|Q_{\cal C}| \geq 1$ as multi-merons. For example, a double-meron
has topological charge $|Q_{\cal C}| = 1$ because it covers a hemisphere of
$S^{N-1}$ (on one side of the reflection hyperplane) twice. In
particular, unlike an instanton, a double-meron does not entirely cover
$S^{N-1}$. An instanton, or any other configuration with $Q = 1$, necessarily 
decomposes
into at least two clusters with half-odd-integer charge. In contrast to 
instantons, which are semi-classical objects that are correlated with each 
other, merons and multi-merons are mutually {\em independent} physical 
topological charge carriers that are well-defined in the full functional 
integral beyond any semi-classical approximation. Via improved estimators they 
are directly related to physical observables and can thus be identified as the 
relevant topological degrees of freedom.

In particular, let us now consider the quantity $\la Q^2 \ra$ that defines the 
topological susceptibility $\chi_t = \la Q^2 \ra/V$, where $V$ is the 
space-time volume. We obtain the improved estimator
\be
\la Q^2 \ra = \left\langle \left(\sum_{\cal C} Q_{\cal C}\right)^2 
\right\rangle =
\left\langle \sum_{\cal C,\cal C'} Q_{\cal C} Q_{\cal C'} \right\rangle =
\left\langle \sum_{\cal C} Q_{\cal C}^2 \right\rangle \ .
\label{meronQ2}
\ee
Since the topological charge $Q_{\cal C}$ changes sign under
cluster flip, and the clusters are independent, $\la Q_{\cal C} Q_{\cal C'} \ra$ 
vanishes for different clusters ${\cal C}$ and ${\cal C'}$.

%% file: 1dO2.tex
The 1-d O(2) model represents a quantum rotor, {\it i.e.}\ a quantum mechanical 
point particle moving on the circle $S^1$, as a Euclidean time path integral.
The model has a topological charge $Q \in \Pi_1[S^1]$ which counts the number
of times the particle moves around the circle $S^1$ during its Euclidean time
evolution. In this case, the space-time volume $V$ is just the Euclidean time 
extent which determines the inverse temperature. In the continuum limit at zero-temperature, 
$V \rightarrow \infty$, one obtains
$\chi_{\rm t} \xi = 1/2 \pi^2$ (where $\xi = 2/g^2$ is the correlation length) \cite{rot97}.
Without any need for a constraint angle $\delta$, the Wolff algorithm builds 
meron-clusters (with $|Q| = 1/2$) and non-meron-clusters (with 
$Q = 0$). As a peculiarity of this simple model, there are no multi-meron 
clusters (with $|Q| \geq 1$) \cite{Boyer}. While it is obvious that the model itself is 
analytically solvable, we will now show that even the properties of the 
corresponding clusters can be derived analytically without any numerical 
simulation. In particular, in this way we obtain the cluster-size distributions 
of meron- and non-meron-clusters.

For the analytical calculation, it is favorable to use the Villain action, which is a {\em quantum perfect action} in this case \cite{rot97}. We then
express $\vec e_x = (\cos\varphi_x,\sin\varphi_x)$ by an angle 
$\varphi_x \in ]-\pi,\pi]$ associated with each lattice site. A pair of 
neighboring spins contributes to the Boltzmann weight according to
\be
\exp\left(- s\left(\varphi_x,\varphi_{x+1}\right)\right) =
\underset{n\in\mathbb{Z}}{\sum}
\exp\left(- \frac{1}{2 g^2} 
\left(\varphi_{x+1} - \varphi_x - 2 \pi n\right)^2\right).
\ee
This weight can be distributed onto configurations with or without a bond, 
according to eq.\ (\ref{bondprob}). Considering the $\varphi = 0, \pi$ axis as 
the reflection hyperplane (in this case, a line), a spin flip sends
$\varphi_x$ to $- \varphi_x$. We define $w_0(\varphi_x,\varphi_{x+1}) = 
\exp\left(- s(|\varphi_x|,- |\varphi_{x+1}|)\right)$ and 
$w_1(\varphi_x,\varphi_{x+1}) = 
\exp\left(- s(\varphi_x,\varphi_{x+1})\right) - w_0(\varphi_x,\varphi_{x+1})$. 
The Wolff cluster algorithm then puts bonds between neighboring spins with 
probability $w_1/(w_0 + w_1)$.

For a system with $V$ sites, a general bond configuration 
$\{\sigma_{12},\sigma_{23},\dots,\sigma_{V1}\}$ is characterized by 
$\sigma_{x,x+1} = 1$ if the bond between the sites $x$ and $x+1$ is put, and
$\sigma_{x,x+1} = 0$ otherwise. This bond configuration occurs with the 
probability $\mathrm{Tr}\left(w_{\sigma_{12}}\dots w_{\sigma_{V1}}\right)/Z$. The 
$w_{\sigma}$ are transfer matrices with elements $w_{\sigma}(\varphi,\varphi\,')$.
The partition function is $Z = \mathrm{Tr}\left(w^V\right)$ with 
$w = w_0 + w_1$.

A cluster with $|{\cal C}|$ sites is characterized by a string of 
$|{\cal C}| - 1$ subsequent matrices $w_1$ between two factors of $w_0$, as 
long as it does not fill the entire volume. Using the cyclicity of the trace, 
the cluster-size distribution of such clusters is given by
\be
\label{csd_all}
p(|{\cal C}|)= \frac{V}{Z} \mathrm{Tr}\left(w_0 w_1^{|{\cal C}|-1} w_0 
w^{V-|{\cal C}|-1}\right).
\ee
Here the factor $V$ accounts for all possible locations of the cluster boundary.
If all sites belong to the same cluster, {\it i.e.} $|{\cal C}| = V$, the value 
of the distribution is 
$p(V) = \mathrm{Tr}\left(V w_0 w_1^{V-1} + w_1^V\right)/Z$. These expressions do
not discriminate between cluster charges.

In order to distinguish the size-distributions of charged meron-clusters 
$p^c(|{\cal C}|)$ from the one of neutral non-meron-clusters $p^0(|{\cal C}|)$
(excluding volume-filling clusters), we construct
\bea
\label{csd_charge}
p^0(|{\cal C}|) = \frac{V}{Z} 
\mathrm{Tr}\left(\left(w_{0+} w_1^{|{\cal C}|-1} w_{0+}+w_{0-} 
w_1^{|{\cal C}|-1} w_{0-}\right) w^{V-|{\cal C}|-1}\right),
\nn \\
p^c(|{\cal C}|) = \frac{V}{Z} 
\mathrm{Tr}\left(\left(w_{0-} w_1^{|{\cal C}|-1} w_{0+}+w_{0+} 
w_1^{|{\cal C}|-1} w_{0-}\right) w^{V-|{\cal C}|-1}\right).
\eea
Here $w_{0\pm}(\varphi,\varphi\,') = w_0(\varphi,\varphi\,')
\theta(\pm(\pi-|\varphi|-|\varphi\,'|))$ characterizes the topology at the 
cluster boundary. Volume-filling clusters can only have vanishing topological 
charge, such that $p^0(V) = p(V)$.

\begin{figure}
\label{exact1dO2}
\vspace{-3mm}
\begin{centering}
\includegraphics[scale=0.48]{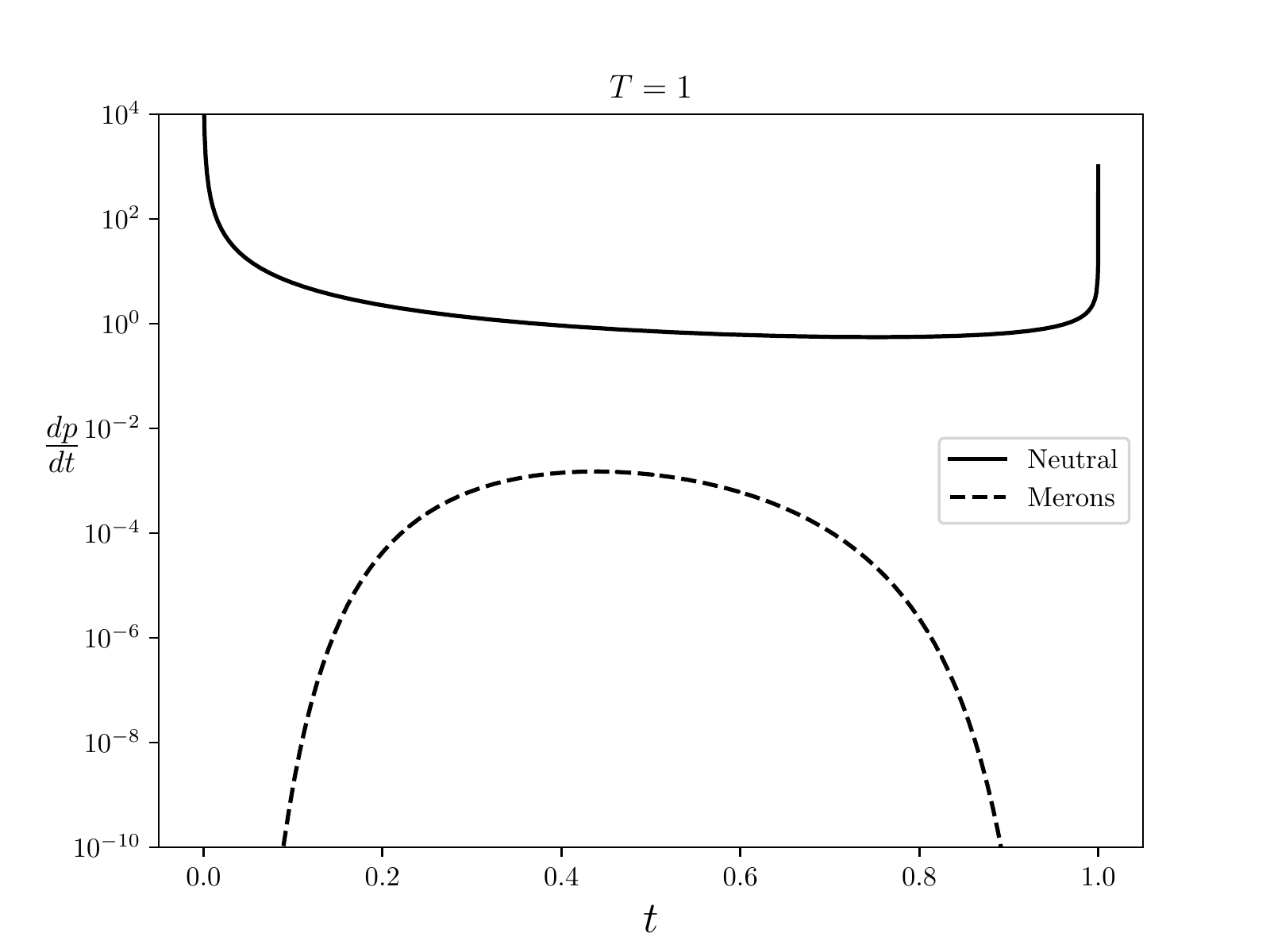}
\includegraphics[scale=0.48]{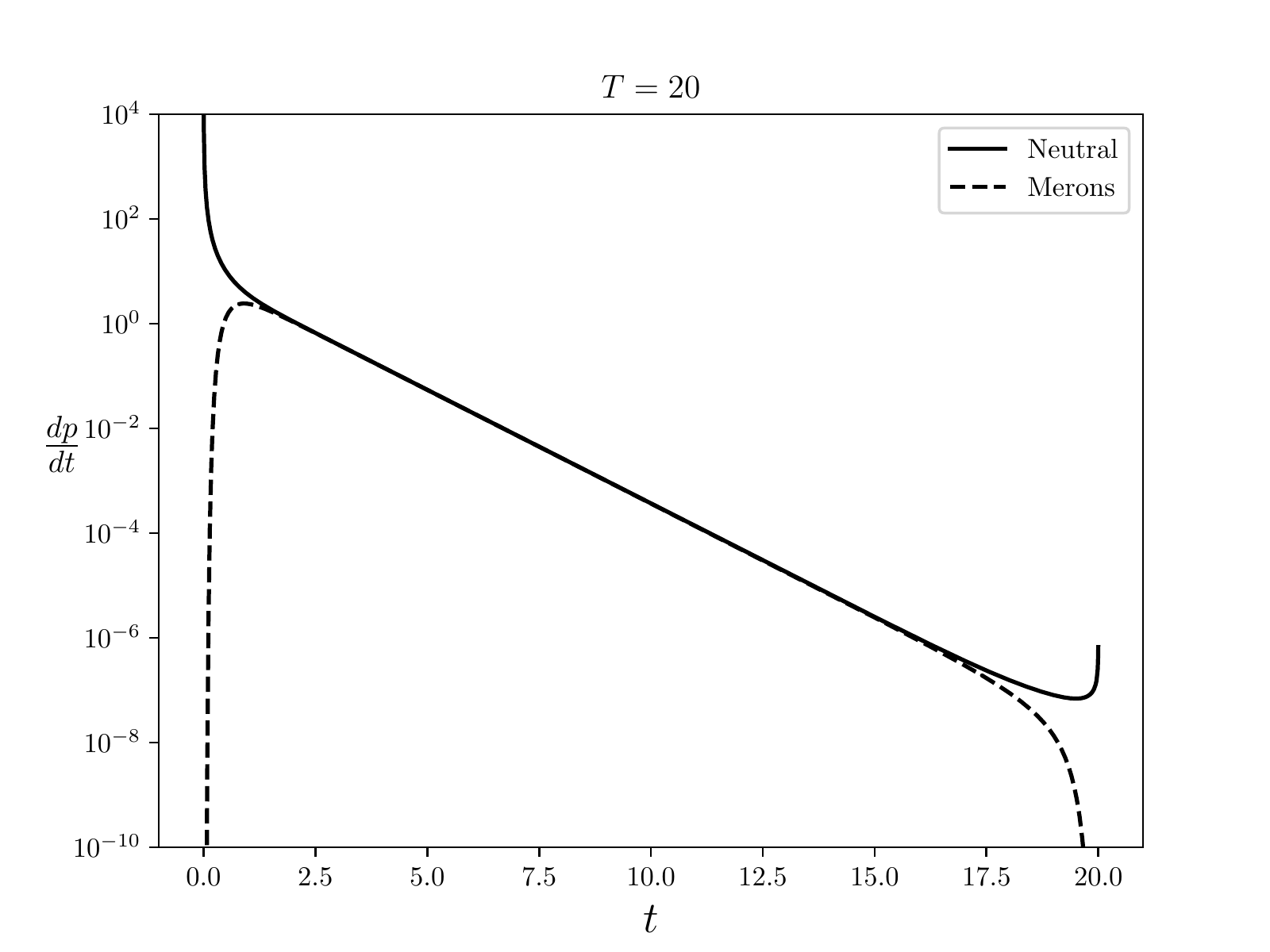}
\end{centering}
\vspace{-6mm}
\caption{Exact cluster-size distribution for the 1-d O(2) model. The plots 
correspond to $T = V/\xi = 1$ and $20$, respectively. For small $T$ 
non-merons are much more abundant than merons.}
\end{figure}

Next, one diagonalizes the various transfer matrices. Finally, one can 
analytically take the continuum limit and express the result in terms of 
Jacobi $\vartheta$-functions
$\vartheta_3(x) = \sum_{n\in\mathbb{Z}} x^{n^2}$ and 
$\vartheta_4(x) = \sum_{n\in\mathbb{Z}}(-1)^n x^{n^2}$.
Defining the functions $z(\lambda) = \vartheta_3(e^{-\lambda})$ and
${\bar z}(\lambda) = \vartheta_4(e^{-\lambda})$, and putting $T = V/\xi$, 
$t = |{\cal C}|/\xi$, we obtain \cite{Stephan}
\be
\frac{dp^0}{dt} = - \frac{2T}{\pi^2} \frac{z(T-t) z'(t)}{z(T)}
+ \frac{z(T) - 1}{z(T)} \delta(t-T), \quad
\frac{dp^c}{dt} = \frac{2T}{\pi^2} \frac{\bar{z}(T-t) \bar{z}'(t)}{z(T)}.
\ee
The $\delta(t-T)$-function results from clusters that fill the entire 
volume. The distributions are illustrated in Fig.\ \ref{exact1dO2} for 
different values of $T$.

%% file: 2dO3.tex
The 2-d O(3) model is asymptotically free, it has a dynamically generated mass 
gap, as well as non-trivial $\theta$-vacua, and thus it shares several
features with QCD. In particular, its topological susceptibility 
$\chi_{\rm t}$ has attracted much attention. Naively, one might expect the 
dimensionless quantity $\chi_{\rm t} \xi^2$ to scale to a finite continuum limit.
However, already a semi-classical analytic calculation in an asymptotically 
small spherical space-time volume reveals a logarithmic ultraviolet divergence 
of $\chi_{\rm t}$ due to a proliferation of instantons of very small size 
$\rho$ \cite{BergLuscher,Luscher82}. The functional determinant of 
perturbative quantum fluctuations around the instanton solution breaks 
classical scale invariance and gives rise to a logarithmically divergent 
contribution to $\chi_{\rm t}$ that is proportional to $\int d\rho/\rho$.

When regularized on the lattice, topological excitations at the lattice spacing
scale --- so-called dislocations, which are lattice artifacts --- may even give 
rise to a power-law divergence of $\chi_{\rm t}$. This has been argued based 
on (non-rigorous) semi-classical considerations \cite{Luscher82}. Ref.\ 
\cite{Blatter} employed a (truncated) classically perfect lattice action, which 
has very small lattice artifacts, and found that dislocations are not present, 
while the logarithmic divergence persists. Recently this property was confirmed 
by applying the gradient flow to the standard action \cite{2dO3GF}. Remarkably, 
the same feature was also observed by simulating the topological action 
\cite{topact}, for which the semi-classical argument would have 
suggested a strong power-law divergence. As a result, dislocation lattice 
artifacts do not seem to be the cause of the ultraviolet divergence of 
$\chi_{\rm t}$. In any case, such a divergence does not imply that topology in 
the 2-d O(3) model is ill-defined in general. For example, the correlation function of the topological charge 
density $\la q_x q_y \ra$ for $x \neq y$ \cite{Balog,topact}, the ratio 
$c_4/\chi_{\rm t}$ (where $c_4 = (3 \la Q^2 \ra^2 - \la Q^4 \ra)/V$ is 
the kurtosis) \cite{Lat16}, as well as the 
$\theta$-dependence of the mass gap \cite{Bogli} are found to converge in the 
continuum limit.

\begin{figure}
\begin{center}
\raisebox{6.5mm}{\includegraphics[angle=0,height=.31\linewidth]{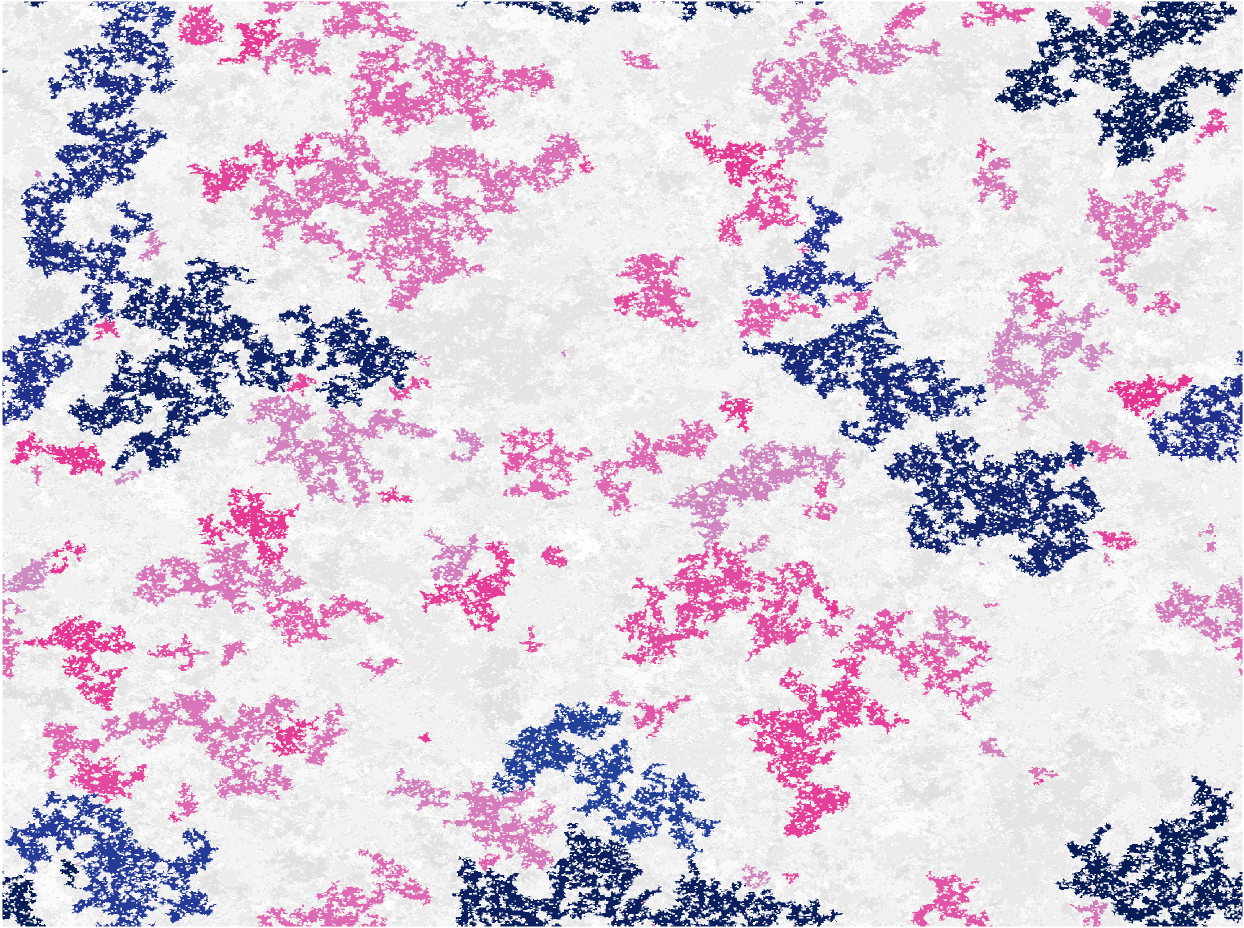}}
\includegraphics[angle=0,height=.38\linewidth]{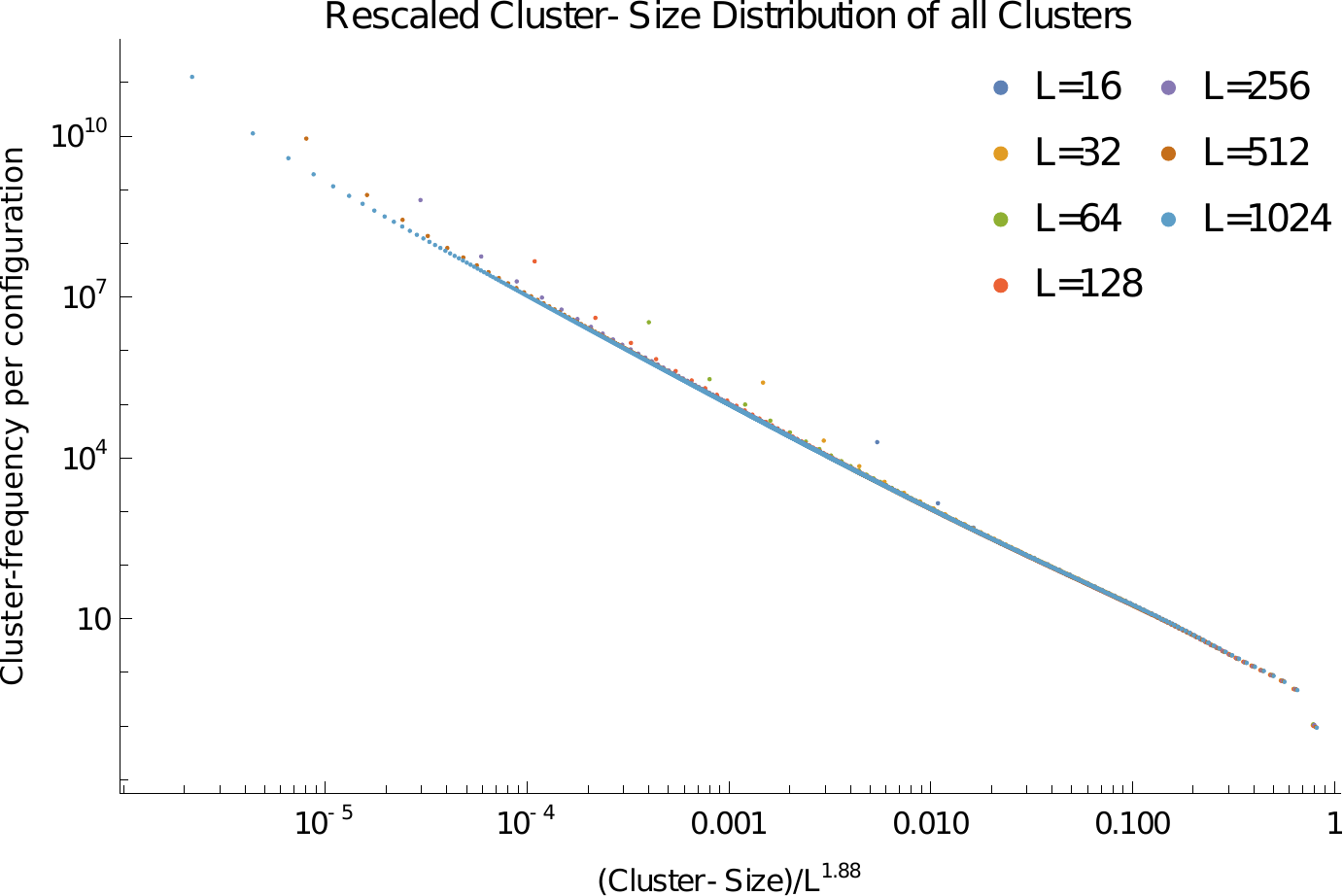}
\end{center}
\vspace*{-3mm}
\caption{a) Left: A typical configuration on a large triangular lattice 
(wrapped around a torus with $L = 1000$) close to the continuum limit. Neutral clusters are shown in light 
gray, while meron- and multi-meron-clusters are darker. b) Right: Scaling of 
the cluster-size distribution for the 2-d O(3) model with the constraint action 
on a triangular lattice, at seven different lattice spacings. Except for the
few-point clusters, the different distributions collapse on a universal 
continuum curve for the fractal dimension $D \simeq 1.88(1)$.}
\label{2dO3clusters}
\end{figure}

Fig.\ \ref{2dO3clusters}a illustrates a typical configuration on a large 
triangular lattice (wrapped around a torus with $L = 1000$) close to the continuum limit, showing a large 
number of meron- and non-meron clusters. Fig.\ \ref{2dO3clusters}b shows the 
scaling of the cluster-size distribution of all clusters. We present data 
obtained on a triangular lattice with the constraint action, for seven 
different lattice spacings, keeping the volume fixed in physical units, {\it i.e.}\
in units of the correlation length. In this case, the system has
the shape of a regular hexagon with sidelength $L = 4 \xi$. 
The different curves collapse on a universal continuum result, provided the 
cluster-size $|{\cal C}|$ is rescaled in units of $L^D$, with the fractal 
dimension $D = 1.88(1)$. The data scale very accurately, except for the tiny 
few-site clusters which represent lattice artifacts that deviate from the 
scaling curve. As a consequence of the reduced fractal dimension $D < d = 2$, 
despite the fact that they are responsible for the long-distance physics,
large clusters only fill a vanishing fraction of space-time in the continuum 
limit. Instead, space-time is filled by few-site clusters which are lattice 
artifacts that are too primitive to develop a fractal structure. Interestingly lower-dimensional topological filaments have also been observed in non-Abelian gauge theories \cite{Horvath}.

\begin{figure}
\vspace*{1mm}
\begin{center}
\includegraphics[angle=0,width=.49\linewidth]{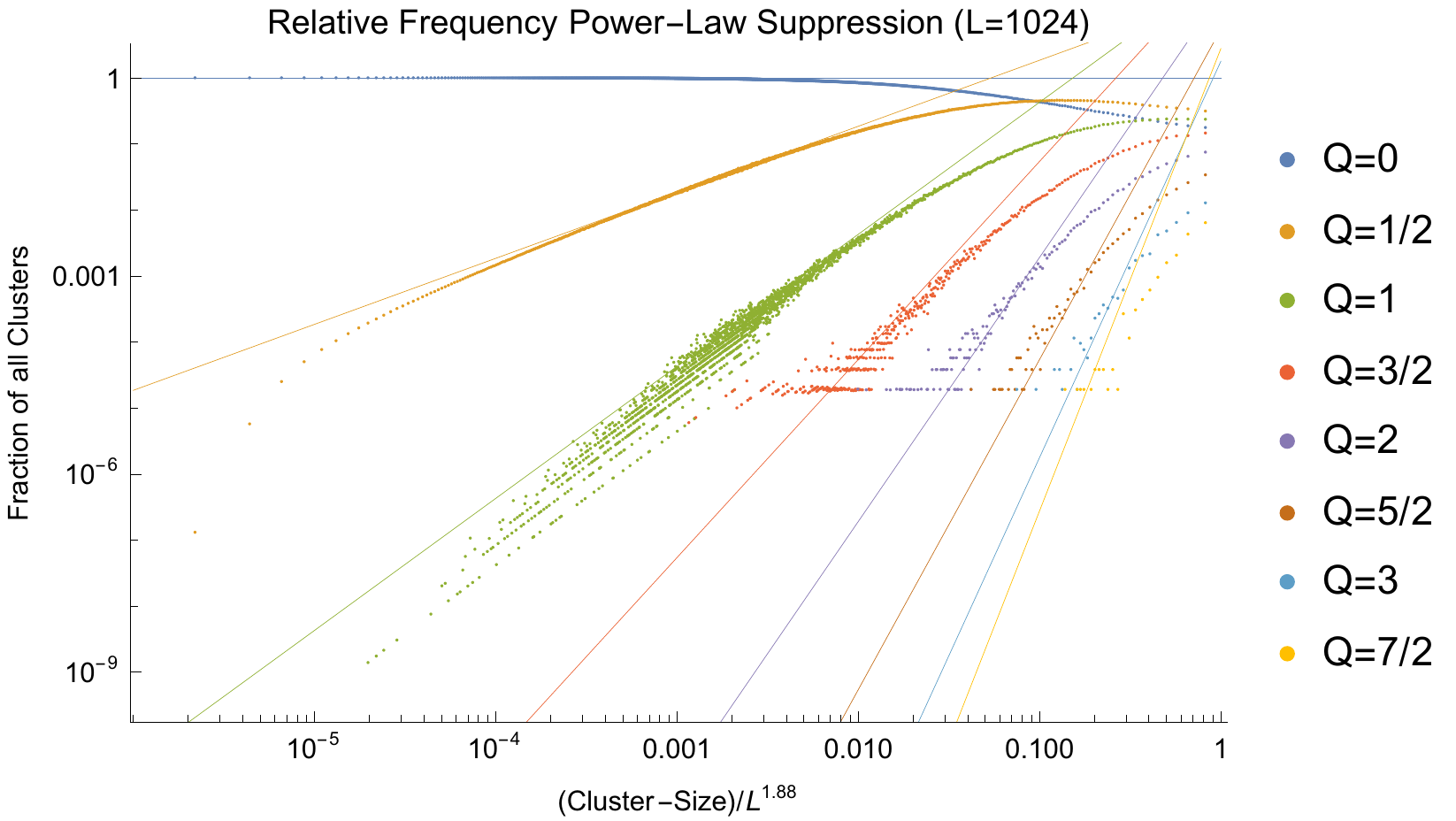}
\includegraphics[angle=0,width=.49\linewidth]{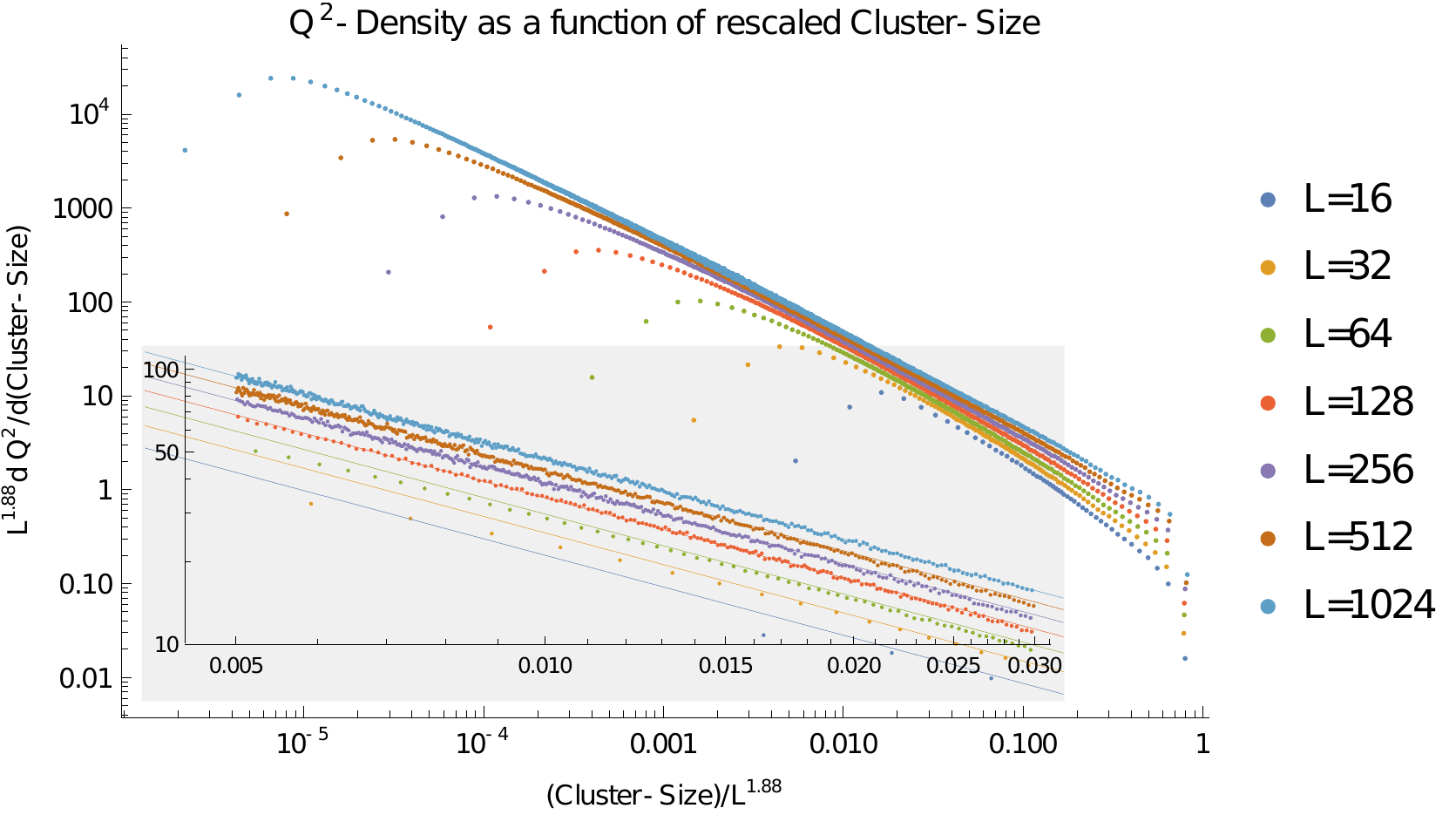}
\end{center}
\vspace*{-3mm}
\caption{a) Left: Fraction of all clusters of a given rescaled cluster-size 
$|{\cal C}|/L^{1.88}$ that carry a topological charge of magnitude $|Q_{\cal C}|$ 
(on the finest triangular lattice with $L = 1024$) and its power-law behavior 
$|{\cal C}|^{- 2 |Q_{\cal C}|}$ (straight lines). b) Right: 
$Q_{\cal C}^2$-contributions of all meron- and multi-meron-clusters of 
a given rescaled cluster-size $|{\cal C}|/L^{1.88}$, for seven different lattice
spacings on a triangular lattice. The inset zooms in on rescaled
cluster-sizes $|{\cal C}|/L^{1.88} \in [0.005,0.03]$ and shows that the
curves do not collapse in the continuum limit.}
\label{2dO3diverge}
\end{figure}

Fig.\ \ref{2dO3diverge}a shows the fraction of all clusters of a given rescaled
cluster-size $|{\cal C}|/L^{1.88}$ that carry a topological charge of magnitude
$|Q_{\cal C}|$. The smallest clusters are non-merons, while multi-merons arise 
only with larger cluster-size. The fraction of clusters with topological charge 
$Q_{\cal C}$ decays as a power-law $|{\cal C}|^{- 2 |Q_{\cal C}|}$. For small 
cluster-size, the predominant contributions to $\chi_{\rm t}$ are due to merons. 
We find an abundance of small
meron-clusters proportional to $1/|{\cal C}|$. Before considering clusters of 
higher charges, $\la Q^2 \ra$ therefore diverges logarithmically, according to 
the integral over the meron distribution. The dimensionless combination 
$\chi_{\rm t} \xi^2 = \la Q^2 \ra (\xi/L)^2$ diverges in the same manner, since 
we keep the physical size $L/\xi $ constant.

In contrast to the 1-d O(2) model, where both merons and non-merons have a 
cluster-size distribution that scales in the continuum limit, it turns out that
in the 2-d O(3) model only the cluster-size distribution of all clusters, but
not the individual distributions of clusters with topological charge 
$Q_{\cal C}$, have a well-defined continuum limit. Fig.\ \ref{2dO3diverge}b
shows the $Q_{\cal C}^2$-contributions of all meron- and multi-meron-clusters of 
a given rescaled cluster-size $|{\cal C}|/L^{1.88}$, for seven different lattice
spacings on a triangular lattice. The curves do not collapse in the continuum
limit. This follows most clearly from the inset that focuses on rescaled
cluster-sizes $|{\cal C}|/L^{1.88} \in [0.005,0.03]$. On finer and finer 
lattices, the multi-meron contributions to $\chi_{\rm t}$ give rise to an 
additional logarithmic divergence, on top of the one due to small merons. 

Via improved estimators, merons also determine the physics at non-zero vacuum
angle $\theta$. At $\theta = \pi$, configurations that contain clusters with
half-integer values of $Q_{\cal C}$ do not contribute to the functional integral,
because the improved estimator for 
$\exp(i \theta Q) = \prod_{\cal C} \exp(i \theta Q_{\cal C})$ then vanishes. 
Configurations with exactly two half-integer clusters still contribute to
$\chi_{\rm t}$, which now results from the improved estimator \cite{meron}
\begin{equation}
\langle Q^2 \exp(i \theta Q) \rangle = \langle Q^2 (-1)^Q \rangle = 
\left\langle \sum_{\cal C} Q_{\cal C}^2 \right\rangle_0 +
2 \langle Q_{{\cal C}_1} Q_{{\cal C}_2} \rangle_2
\end{equation}
The first term on the right-hand side receives contributions from 
configurations that contain no clusters with half-integer $Q_{\cal C}$. The 
second term results from configurations with exactly two clusters of half-integer charges with values $Q_{{\cal C}_1}$ and
$Q_{{\cal C}_2}$ in the reference configuration. Since such 
configurations are much more abundant than those that contribute to the first
term, the second term implies an infrared divergence of $\chi_{\rm t}$ in the 
infinite-volume limit, and thus induces a phase transition at $\theta = \pi$. 
Merons indeed explain the mechanism that is responsible for this transition.

%% file: 2dO2.tex
In the 2-d O(2) model there is no global topological charge. However, on a 
square lattice an integer vorticity $v_\Box \in \{-1,0,1\}$ is associated with 
each plaquette. Two nearest-neighbor spins $\vec e_x$ and $\vec e_y$ 
on a plaquette define a shortest arc on $S^1$. The sum of the four oriented 
arc lengths associated with the four nearest-neighbor pairs of the plaquette,
added up anti-clock-wise, divided by the circumference $2 \pi$ of $S^1$, define 
the vorticity $v_\Box$. Due to periodic boundary conditions, along with Stokes' 
theorem, the vorticity summed over all plaquettes vanishes, 
$\sum_\Box v_\Box = 0$, {\it i.e.}\ each configuration has the same number of 
vortices and anti-vortices.

The 2-d O(2) model has a phase transition of infinite order \cite{BKT1,BKT2}, 
which occurs for the standard action at $1/g_{\rm c}^2 = 1.1199$ \cite{Has05} and
for the topological action at $\delta_{\rm c} = 1.775(1)$ \cite{BBNPRW,BGR13}.
In agreement with the Mermin-Wagner-Coleman theorem, no symmetry breaking 
occurs at this transition and no long-range order is generated. However, the 
transition separates a massive from a massless phase. In the massless phase, 
spin correlations decay algebraically with a critical exponent $\eta$ which 
varies continuously inside this phase. Hence, the massless phase corresponds to 
a family of different universality classes.

The celebrated Berezinski\u{\i}-Kosterlitz-Thouless (BKT) mechanism explains 
this transition as follows: in the massless phase vortices and anti-vortices are
rare and predominantly appear in tightly bound pairs. Beyond the transition 
there is a significant number of vortices and anti-vortices which unbind, thus 
replacing the algebraically decaying correlations by 
exponential decays. The proliferation of unbound vortices and anti-vortices is 
hence responsible for the generation of a mass gap. Remarkably, the same 
mechanism still works for the topological action \cite{BGR13}. Then the action 
vanishes for all allowed configurations, and vortex (un)binding is triggered solely by 
entropy. 

Interestingly, the vortex--anti-vortex dynamics is directly reflected in the
structure of the corresponding Wolff clusters. Similar to the global topological
charge $Q = \sum_{\cal C} Q_{\cal C}$ of the $(N-1)$-dimensional O($N$) model, 
which is a sum of cluster charges, in the 2-d O(2) model the local vorticity 
$v_\Box = \sum_{\cal C} v_{\Box,{\cal C}}$ of a plaquette also receives independent
additive contributions $v_{\Box,{\cal C}}$ from different clusters ${\cal C}$. 
Since the vorticity changes sign under reflection at the hyperplane (in this 
case a line dividing $S^1$ into two semi-circles), in analogy to 
eq.\ (\ref{Qclust}) the contribution of a given cluster ${\cal C}$ to the 
vorticity on a specific plaquette $\Box$ is given by
\be
v_{\Box,{\cal C}} = \tfrac{1}{2} (v_\Box[\vec e \, ] - v_\Box [\vec e \, ']) 
\in \{ 0, \, \pm \tfrac{1}{2}\} \ .
\ee
Again $[\vec e \, ']$ is the configuration that one obtains after flipping the 
cluster ${\cal C}$, while keeping all the other clusters fixed. It should be
noted that, just as for the 1-d O(2) model, it is not necessary to impose a 
constraint angle $\delta$ between nearest-neighbor spins because 
$v_{\Box,{\cal C}}$, as defined above, is already independent of the relative
orientations of all the other clusters.

As before, all spins within a given cluster are on the same side of the
reflection line. Hence, a vortex or anti-vortex (which covers all of $S^1$) 
cannot be contained inside a single cluster. Instead, it is dissected into two 
semi-vortices or semi-anti-vortices that reside in two different clusters.\footnote{It turns out that the vorticity of a plaquette is never shared by 
more than two clusters.} Semi-vortices can be distinguished from 
semi-anti-vortices by flipping the clusters into a reference configuration with 
all spins on the same side of the reflection line. Semi-vortices and 
semi-anti-vortices reside at the cluster boundary in alternating order. One can 
thus classify the clusters by the number of semi-vortex--semi-anti-vortex pairs 
located at the cluster boundary. 

The cluster decomposition sheds new light on the details of the BKT mechanism.
Vortex-anti-vortex pairs become well-defined and tractable via a sub-ensemble of
$2^{n_{\cal{C}}}$ configurations, which allows for direct investigation of
their correlation. We distinguish three types of pairs: uncorrelated,
sign-correlated, and bound. In uncorrelated pairs all four semi-(anti-)vortices 
reside on different clusters, thus a total of four clusters are involved. The 
vorticity of such a pair is completely uncorrelated. Sign-correlated pairs are 
connected by a single cluster, comprising one semi-(anti-)vortex at each 
plaquette of the pair. This fully correlates the sign of the vorticities. The 
existence of a vortex, however, is governed by the two remaining 
semi-(anti-)vortices, which reside in different clusters, thus rendering them 
uncorrelated. Bound pairs, on the other hand, are connected by two clusters, 
each carrying one semi-(anti-)vortex at both plaquettes of the pair. Since this
correlates the sign as well as the existence of the two vortices 
simultaneously, they indeed form a {\em bound pair}. A quantitative 
investigation of this concept of vortex-anti-vortex binding and unbinding is 
currently in progress, and holds the promise to further deepen our 
understanding of the BKT mechanism.

\begin{figure}
\vspace*{1mm}
\begin{center}
\includegraphics[angle=270,width=.49\linewidth]{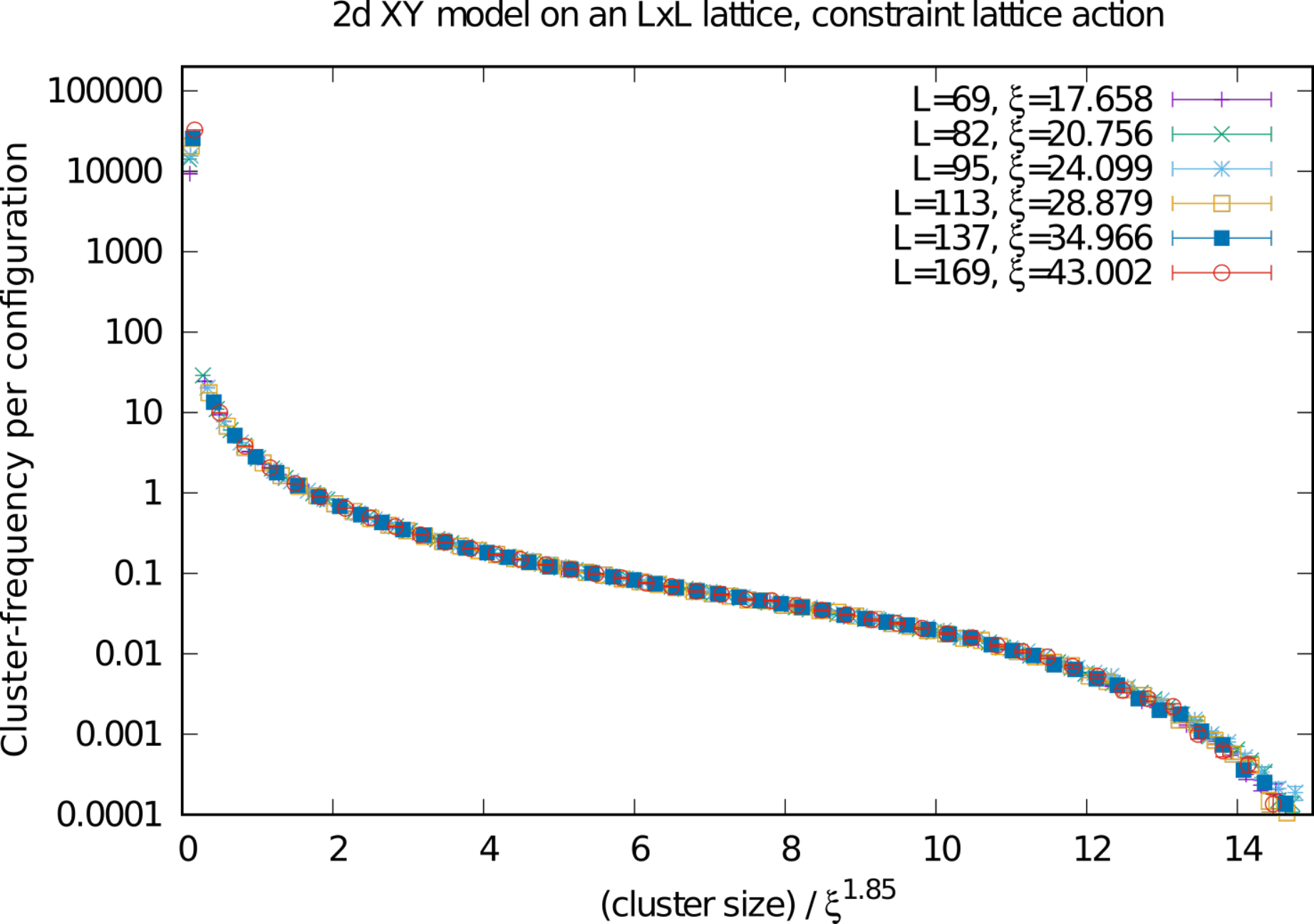}
\hspace*{1mm}
\includegraphics[angle=270,width=.49\linewidth]{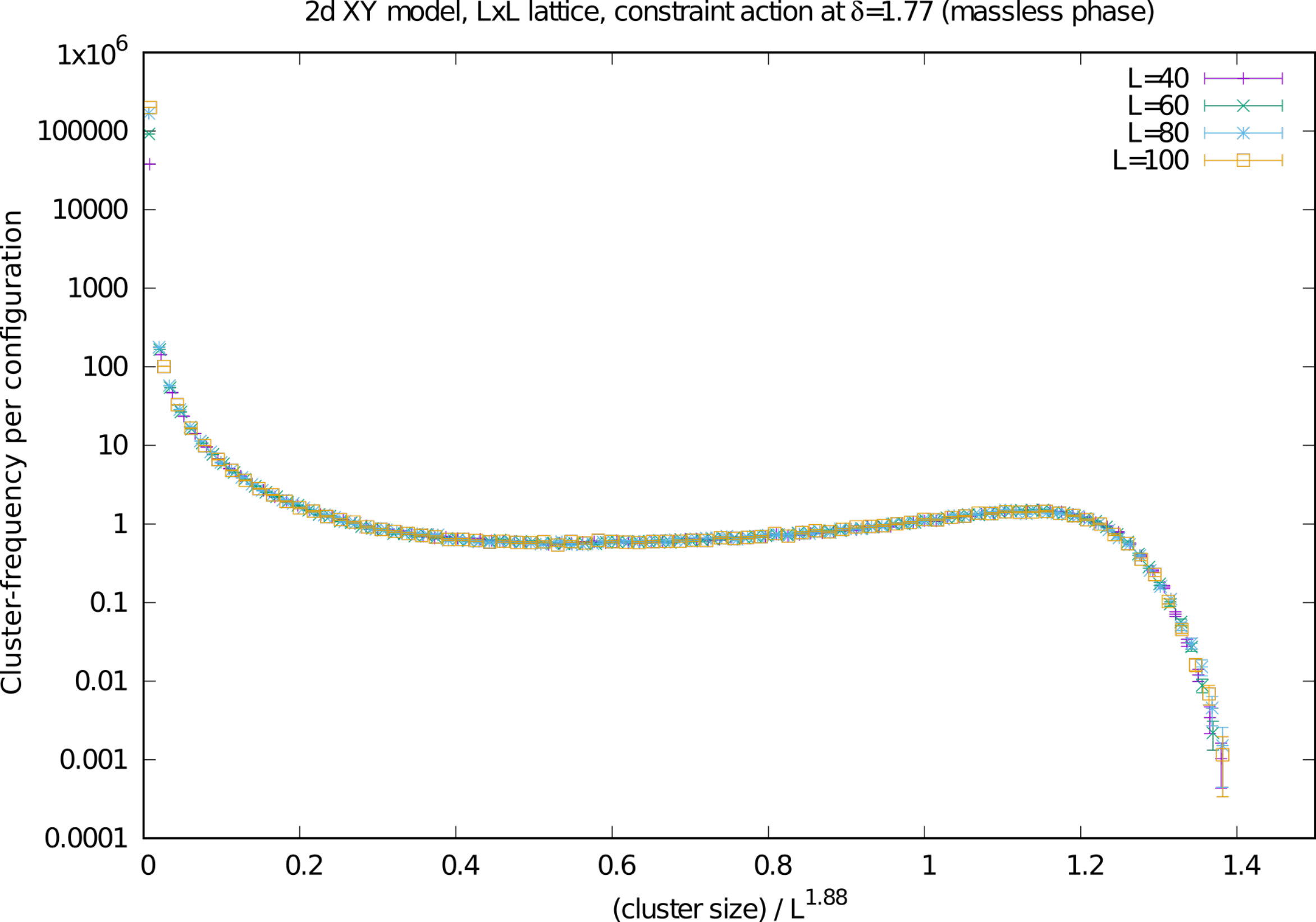}
\end{center}
\vspace*{-3mm}
\caption{Cluster-size scaling for the 2-d O(2) model with the topological 
action for several different lattice spacings. a) Left: In the massive phase, 
at $L/\xi = 3.93(1)$, ignoring few-site clusters, the cluster-size distribution 
scales with fractal dimension $D = 1.85(1)$. b) Right: In the massless phase at
$\delta = 1.77$ (with $L/\xi_2 \simeq 1.34$) one obtains $D = 1.88(1)$.}
\label{XYscaling}
\vspace*{-1mm}
\end{figure}

Now we address the issue of cluster-size scaling, first in the massive phase.
Fig.\ \ref{XYscaling}a shows the cluster-size distributions $p(|{\cal C}|)$ 
obtained for the topological action  for six different lattice spacings, where $\delta$ is tuned in each case such that the physical volume is fixed to $L/\xi = 3.93(1)$; $L$ 
ranges from $69$ to $169$, $\xi$ from $17.7$ to $43.0$, and $\delta$ from $2$ 
to $1.95$. Ignoring few-site clusters, which are lattice artifacts, scaling is 
well confirmed if the cluster-size is expressed in units of $\xi^D$, where 
$D = 1.85(1)$ is the {\em fractal dimension.}

Next, we proceed to the massless phase, {\it i.e.}\ 
$\delta \leq \delta_{\rm c} \simeq 1.775$. Here each value of $\delta$, or each 
$1/g^2 \geq 1/g_{\rm c}^2$ for the standard action, represents its own 
universality class. In a finite volume of size $L \times L$, $\xi$ is finite as 
well, and it turns out that --- at fixed $\delta$ --- the physical ratio 
$L/\xi$ is practically $L$-independent. This also holds for the second moment 
correlation length $\xi_2$ which is very close to $\xi$ but easier to measure 
($\xi_2$ tends to be slightly shorter). For simplicity we are going to refer to 
$\xi_2$; its definition is reproduced {\it e.g.}\ in Ref.\ \cite{topact}, 
Section 4. Fig.\ \ref{XYscaling}b refers to $\delta = 1.77$ and 
$L=40,\dots,100$, where we consistently obtained $L/\xi_2 \simeq 1.34$. Here 
the cluster-size is given in units of $L^D$, with the fractal dimension 
$D = 1.88(1)$, which leads to precise cluster-size scaling (except for the 
few-site clusters).

When we repeat these simulations deeper in the massless phase, the ratio
$L/\xi_2$ shrinks while $D$ gradually increases. For 
$\delta = 1.775, 1.7, 1.6$, and $1.5$, we obtain 
$D = 1.88(1), 1.89(1), 1.91(1)$, and $1.93(1)$, respectively. Ultimately, $D$ 
converges to $2$. Indeed, when $\delta$ approaches 0, all spins must be 
parallel and all bonds between nearest neighbors will be put with probability 
1. As a result, at $\delta = 0$ there is just one cluster which fills the 
entire space-time volume, and thus has dimension $D = 2$.

%% file: 3dO4.tex
The 3-d O(4) model can be interpreted as a high-temperature effective theory 
that captures the universal features of the finite temperature chiral phase 
transition of 2-flavor QCD in the chiral limit, provided that this transition
is second order \cite{Rajagopal}. This is due to the equivalence of the 
symmetry breaking 
pattern ${\rm SU}(2)_L \times {\rm SU}(2)_R \to {\rm SU}(2)_{L=R}$ with
${\rm O}(4) \to {\rm O}(3)$. The 4-d O(4) model has topological Skyrmion 
excitations which may be interpreted as baryons \cite{Skyrme}. At any fixed 
time, the total baryon number $B$ is given by the winding number in the 
homotopy group $\Pi_3(S^3) = \Z$. The fermionic statistics and spin of the 
Skyrme-baryons are determined by the two elements of the homotopy group 
$\Pi_4(S^3) = \Z(2) = \{\pm 1\}$, which represent the fermion sign in the 
corresponding functional integral. When the temperature is sufficiently high to 
imply dimensional reduction, Skyrmions become static and their sign problem 
associated with $\Pi_4(S^3)$ disappears. Their topological baryon number,
$B \in \Pi_3(S^3) = \Z$, however, persists at high temperatures.

In this model, the {\em meron concept}, {\it i.e.}\ the assignment of 
well-defined half-integer topological baryon numbers to individual clusters, 
requires the restrictive constraint $\cos\delta > - 1/3$ on the relative 
angles between neighboring spins within a simplex, such that only rather smooth 
configurations are admitted. It turns out that imposing this constraint 
restricts us to the low-temperature chirally broken phase. Since we are 
interested in the universal features of the chiral phase transition, we 
consider the standard action instead and study the cluster-size scaling for all 
clusters irrespective of their topological features. The critical inverse 
coupling was identified as $1/g_{\rm c}^2 = 0.93590(5)$ \cite{Oevers,EFS}.

\begin{figure}
\vspace*{-2mm}
\begin{center}
\includegraphics[angle=0,width=.5\linewidth]{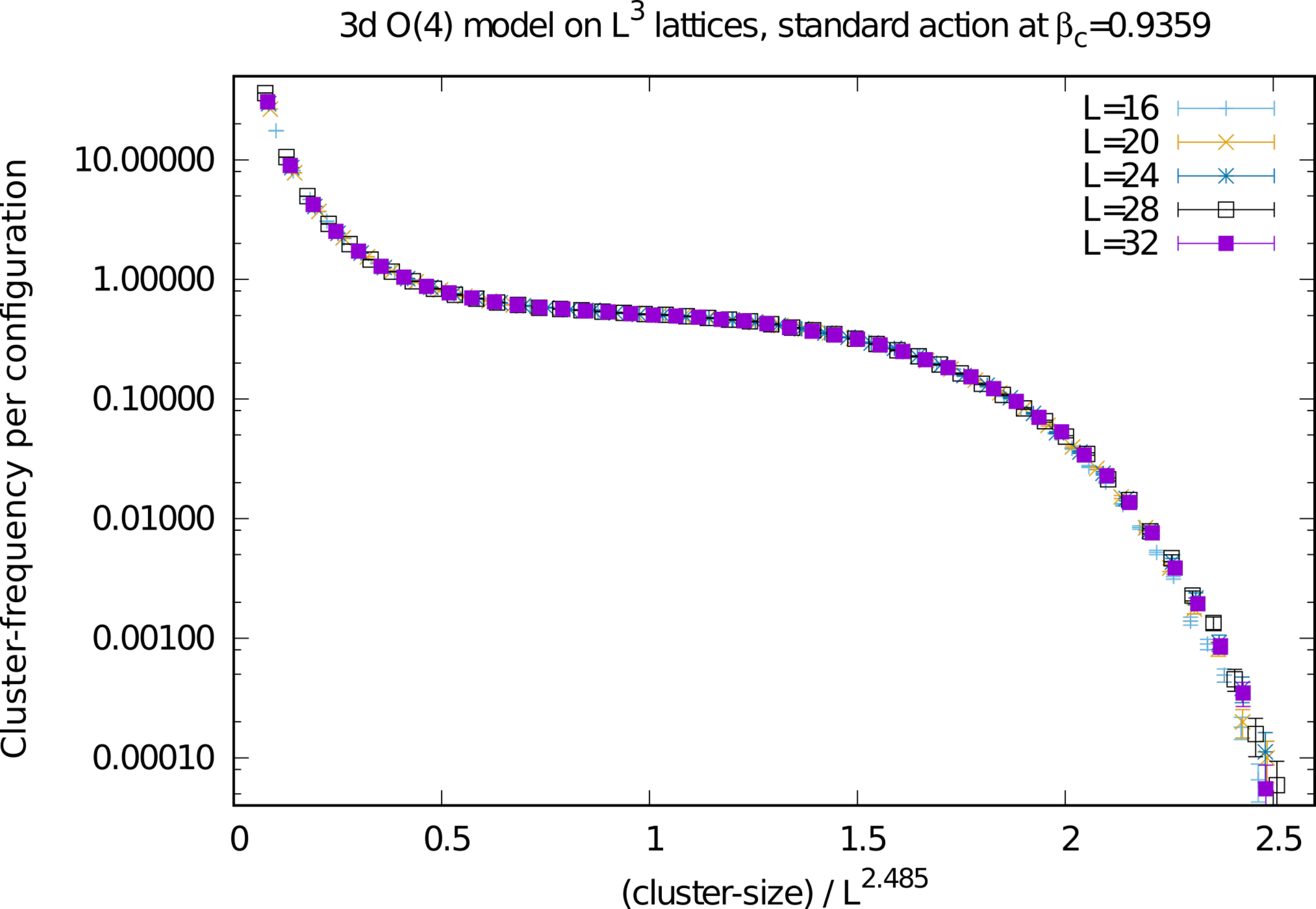}
\hspace*{-4mm}
\includegraphics[angle=0,width=.51\linewidth]{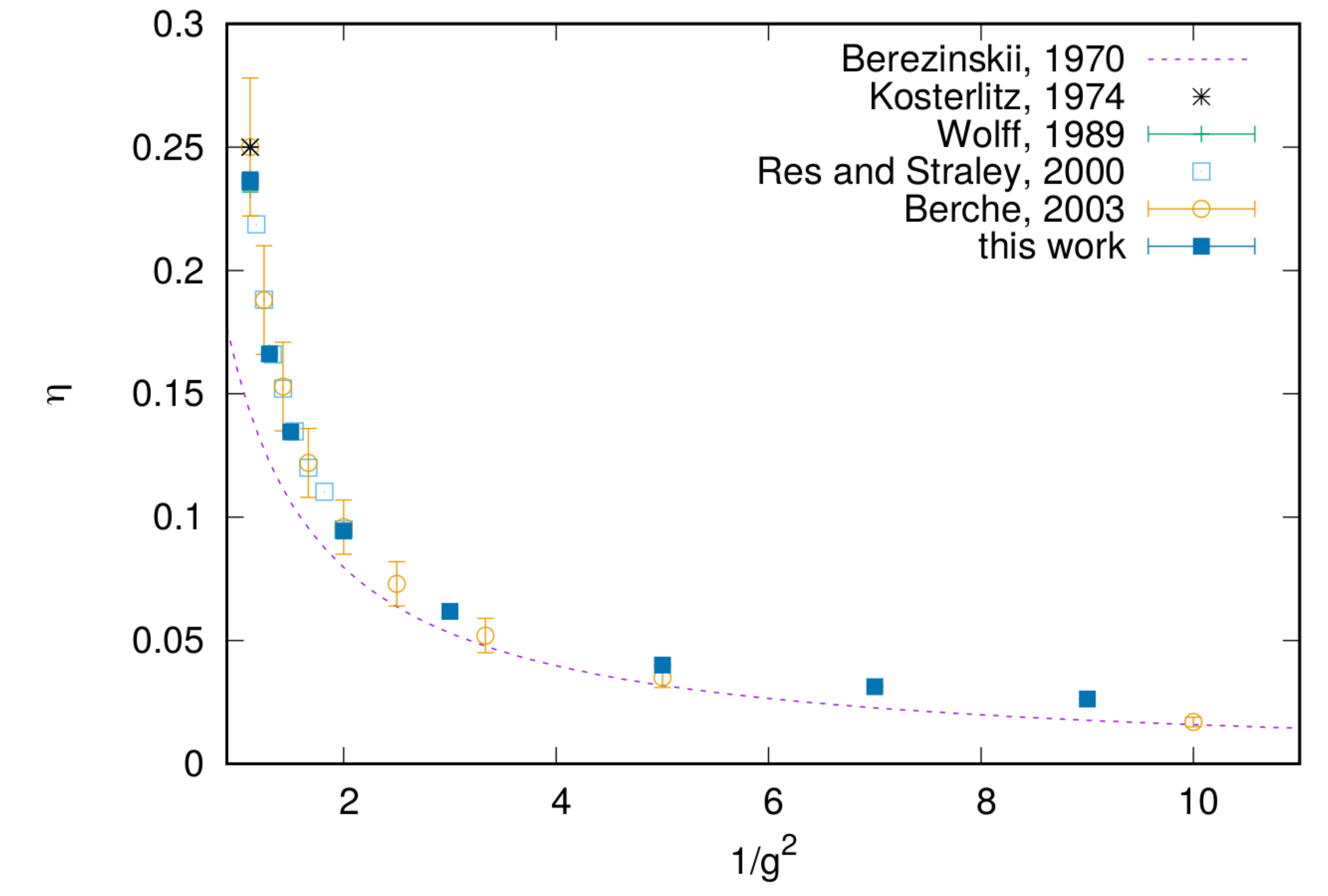}
\end{center}
\vspace*{-4mm}
\caption{a) Left: Cluster-size scaling in the 3-d O(4) model, in volumes $L^3$,
with the standard action at the critical coupling. b) Right: Fractal dimension
$D$ for the 2-d O(2) model in the massless phase, for the topological action.
The values of $\eta = 2(2-D)$ obtained from our measurements of the fractal
dimension $D$ are compared with values of $\eta$ from the literature 
\cite{BKT1,Kost74,Wolff2,Res,Berche}. The dotted line is
the leading order analytic prediction for large values of $1/g^2$.}
\label{3dO4cluscal}
\vspace*{-1mm}
\end{figure}

A cluster-size scaling study is motivated at the critical point.
Fig.\ \ref{3dO4cluscal}a shows the matching of the histograms for
$L^3$ lattices, where the cluster-size is successfully rescaled by
$L^D$ with $D=2.485$. 

It is interesting to ask whether the fractal dimension is
related to the critical exponents of the chiral phase transition. Such a 
relation indeed exists for the $d$-dimensional Ising model. In that
case, the magnetization $M = \sum_{\cal C} M_{\cal C}$ receives additive 
contributions from the individual Swendsen-Wang clusters \cite{SwendsenWang}. 
All Ising spins inside a given cluster are parallel, so the absolute value of 
the magnetization of a cluster $|M_{\cal C}|$ is simply given by the cluster-size
$|{\cal C}|$. Since the cluster magnetization changes sign under cluster flip, 
and since all clusters are independent, one obtains an improved estimator for 
the magnetic susceptibility 
\be
\chi_{\rm m} = \frac{1}{V} \la M^2 \ra = 
\frac{1}{V} \left\langle \left(\sum_{\cal C} M_{\cal C}\right)^2 \right\rangle 
= \frac{1}{V} \left\langle \sum_{\cal C} |{\cal C}|^2 \right\rangle \ .
\ee
Hence, the cluster-size, whose distribution determines the fractal dimension,
is directly related to $\chi_{\rm m}$. Using finite-size scaling of 
$\chi_{\rm m}$, for the Ising model one can analytically relate the fractal 
dimension to the critical exponents $\beta$ and $\nu$ \cite{Stauffer,Janke}
\be  \label{scalinglaw}
D = d - \beta/\nu \ .
\ee

In the O($N$) models, all spins inside a Wolff-cluster are on the same side of 
the reflection hyperplane, but they are typically not parallel. As a result, 
the cluster-size is no longer directly related to $\chi_{\rm m}$ and thus the 
fractal dimension can no longer be analytically related to the critical 
exponents by the same argument. Remarkably, we find that the scaling law 
\eqref{scalinglaw} also works in the 3-d O(4) model to high numerical
accuracy. Ref.\ \cite{EFS} reported (and we have independently confirmed) 
$\beta = 0.380$ and $\nu=0.7377$, which implies $d - \beta / \nu = 2.485$, in 
excellent agreement with our measurement of the fractal dimension $D$.

Encouraged by this result, we apply the scaling law \eqref{scalinglaw} also to 
the 2-d O(2) model in the massless phase. It should be noted that, in that 
case, due to the exponentially diverging correlation length and susceptibility, 
$\beta$ and $\nu$, which characterize power-law divergences, are not defined. 
Based on hyperscaling, $2 \beta/\nu$ is then replaced by the continuously 
varying critical exponent $\eta \leq \eta_{\rm c} = 1/4$ \cite{Kost74}, which 
characterizes the algebraic decay of spin correlations in the massless phase. 
Thus, $\eta$ determines the finite-size scaling of the susceptibility in the 
large-volume limit, $\chi_{\rm m} \propto L^{2 - \eta}$.
Replacing $\beta/\nu$ by $\eta/2$ in the scaling law \eqref{scalinglaw}, we
obtain $D = 2 - \eta/2$. Fig.\ \ref{3dO4cluscal}b shows $D$ as measured 
at various values of 
$1/g^2 > 1/g_{\rm c}^2$ in comparison with $d - \eta/2$. Indeed, the scaling law
also holds in the 2-d O(2) model both at the BKT transition and inside the 
massless phase. The same has been observed using the topological action. At 
the BKT transition, $\eta$ approaches $\eta_{\rm c} = 1/4$ which implies 
$D = 1.875$, in agreement with the observed fractal dimension $D = 1.88(1)$.
In view of the scaling law, it would be interesting to also understand the
value of the fractal dimension, again $D = 1.88(1)$, that we obtained for the 
2-d O(3) model.

%% file: conclu.tex
Based on the cluster-size continuum scaling that we observed in a set of O($N$) 
models, we have argued that clusters can be interpreted as physical objects. 

In the quantum mechanical 1-d O(2) model, the dimension $D$ of the clusters 
coincides with the dimension $d=1$. For $d$-dimensional quantum field theories 
with $d>1$, however, we observed a fractal dimension $D < d$. This implies that 
the physical clusters fill only a negligible fraction of space-time. In the 
continuum limit, space-time is filled with tiny few-site clusters, which do not 
display this scaling behavior. We also observed this in the broken phase of the 
3-d O(4) model, where $D$ approaches $3$ as $1/g^2$ increases. In the Ising 
model the analytically derived scaling law (\ref{scalinglaw}) relates the 
fractal dimension $D$ to the critical exponents. We postulated the 
applicability of this scaling-law for general O($N$) models, and confirmed this 
numerically in the 2-d O(2) and 3-d O(4) model.

In the 2-d O(2) model, we identified semi-vortices and semi-anti-vortices that
reside at the cluster boundaries in alternating order. Based on this, we
introduced a criterion to decide whether a vortex and an anti-vortex form a
bound pair. This holds the promise to provide deeper qualitative and 
quantitative insights into the mechanism of the BKT phase transition.

In models with a global topological charge, if an appropriate constraint on the
relative angle between nearest-neighbor spins is implemented in the lattice 
action, we assigned a uniquely defined topological charge $Q_{\cal C}$ to each 
cluster. In particular, the merons (with $|Q_{\cal C}| = 1/2$) are important 
carriers of topological charge. In the 1-d O(2) model, merons persist in the
continuum limit, as we even showed analytically. In the 2-d O(3) model, on the 
other hand, the situation is more subtle. First of all, a power-law ultraviolet
divergence of the topological susceptibility $\chi_{\rm t}$ due to dislocation 
lattice artifacts does not seem to arise. Instead, just like small instantons 
in a semi-classical investigation, meron-clusters of a small physical size give 
rise to a logarithmic divergence of $\chi_{\rm t}$. However, while the 
cluster-size distribution of all clusters collapses on a universal continuum 
curve, the size-distributions of the clusters with fixed topological charge 
$Q_{\cal C}$ do not converge in the continuum limit. In fact, 
multi-meron-clusters proliferate, diminish the probability of meron- and
non-meron-clusters, and give rise to yet another logarithmic divergence of 
$\chi_{\rm t}$.

It would be most interesting to extend the stochastic definition of the 
physical topological charge carriers to other models, including 2-d $\CP(N-1)$ 
models for $N>2$, the Schwinger model, as well as 4-d Yang-Mills theories, or 
QCD. In particular, if these objects could be tied directly to physical
observables, via something like improved estimators, they could be established 
as relevant physical degrees of freedom with a solid field theoretical basis, 
beyond any semi-classical approximation. Unfortunately, for the models mentioned
above, efficient cluster algorithms could not be constructed until now. 
In particular, Wolff-type embedding cluster algorithms are not efficient in 
$\CP(N-1)$ models with $N>2$ \cite{JansenWiese,Sokal}, and do not give rise to 
a meaningful definition of merons. Our study provides a strong motivation to
keep searching for appropriate constructions, in order to deepen our 
understanding of the topological mechanisms that are responsible for the 
generation of a mass gap in non-linear $\sigma$-models and for confinement in 
gauge theories.